\documentclass[%
 aip,%
 amsmath,amssymb,
 unsortedaddress,
 reprint,%
]{revtex4-1}

\usepackage{graphicx}
\usepackage{dcolumn}
\usepackage{bm}

\begin{document}

\title[Learning from the dynamics of the Covid-19 epidemic]
{What can we learn from the dynamics of the Covid-19 epidemic ?}

\author{M. Peyrard}
\email{Michel.Peyrard@ens-lyon.fr}
\affiliation{%
Ecole Normale Sup{\'e}rieure de Lyon,
Laboratoire de Physique CNRS UMR 5672, 46 all{\'e}e d'Italie, F-69364 Lyon
Cedex 7, France}

\date{\today}

\begin{abstract}
  We investigate the mechanisms behind the quasi-periodic outbursts on the
  Covid-19 epidemics. Data for France and Germany show that the patterns of
  outbursts exhibit a qualitative change in early 2022, which appears in a
  change in their average period, and which is confirmed by the time-frequency
  analysis. This provides a signal that can be used to discriminate among
  several mechanisms. Two main ideas have been proposed to explain periodicity
  in epidemics. One involves memory effects and another considers exchanges
  between epidemic clusters and a reservoir of population. We test these two
  approaches in the particular case of the Covid-19 epidemics and show that
  the ``cluster model'' is the only one that appears to be able to explain the
  observed pattern with realistic parameters. The last section discusses our
  results in the context of early studies of epidemics, and we stress the
  importance to work with models with a limited number of parameters, which
  moreover can be sufficiently well estimated, to draw conclusions on the
  general mechanisms behind the observations.
\end{abstract}

\maketitle

\begin{quotation}
The spreading of the Covid-19 epidemic over the world has introduced chaos in
the life of many people and in the international economy. But besides this
familiar concept of ``chaos'', theoretical studies have shown that the epidemic
can be formally characterized as a chaotic dynamical system in the
mathematical sense \cite{Jones2020}. However chaos and order do not exclude
each other. In the preface of the proceedings of the International Conference
on ``Order in Chaos'', held in the Center for Nonlinear Studies of Los Alamos
in 1982, David Campbell and Harvey Rose pointed out that, counter to
intuition, deterministic systems can exhibit a chaotic behavior
\cite{Campbell1983}. In the case of the Covid-19 epidemic, the question which
arises is the opposite one: although it appears as truly complex and
chaotic, can a system exhibit orderly patterns which can be related to a
simple underlying mechanism ? And if it is so, can we learn something on this
mechanism by analyzing these patterns ? Figure~\ref{fig:covidcases}
displaying the time evolution of the number of daily Covid-19 new cases
during the development of the epidemic in France and Germany from early 2020 to
May 2023 shows a quasi-periodic recurrence of outbreaks which suggests some
underlying regularity. As the epidemic expanded, it was often mentioned that
the disease would be characterized by seasonal outbreaks that could easily be
understood because in cold seasons people tend to gather indoors, which
increases the risk of contamination. However the period of the oscillations
observed in Fig.~\ref{fig:covidcases} is of the order of three months, showing
that this naive explanation of the periodicity is wrong. There are outbreaks
in all seasons, in spring as well as in winter, in both France and
Germany. Another suggestion could be the dynamics of social life, with
vacations alternating with work periods. Again, although it might play some
role, the data show that this explanation is incomplete. This precludes naive
explanations but this may actually be extremely useful to open a path to
understand some fundamental properties of the Covid-19 epidemic by an analysis
of the data. This is what this paper intends to show.

\end{quotation}

\section{Introduction}
\label{sec:intro}

A bibliographical search for the investigations of the spreading of the
Covid-19 epidemic finds thousands of papers.
Some of them rely on a statistical
analysis focusing on medical aspects such as the mechanisms of the
person-to-person transmission or the effectiveness of vaccines. Many studies
are devoted to the modeling of the epidemic and they build on a long history of
research, including pioneering studies, in which 
many of them devoted to the measles
epidemics in England \cite{Kermack1927, Soper1929,Bartlett1957}. These papers
introduced the susceptible, infected, recovered (SIR) model which is
the basis on which many modern studies are build.

The objectives of all this research are
diverse. Part of it is used as a guide to make decisions for the containment of
the epidemic. In this case the results have to be as quantitatively accurate
as possible and, therefore take into account various phenomena concerning the
disease, its incubation period and means of transmission, the details of the
contacts between individuals depending on local transportation networks, and
so on. This implies that such models include many parameters that have to be
carefully fitted in each specific situation. These studies may have a great
practical utility but they are not suitable to understand the general
mechanisms behind the epidemics.

Another goal is to look for these basic mechanisms without attempting a
detailed fit of the data but instead trying to determine what is required
in a model to get the main features which are observed in various countries or
locations. The origin of periodic outbursts is one of such general questions,
which attracted a lot of attention since the very early studies
\cite{Soper1929,Bartlett1957}. This question is not trivial. It is tempting to
attribute the periodicity to some external effect, which amounts to
introducing in the model some time-dependent parameters which oscillate in
time. This is how we explain the recurrence of the epidemics of flu that affect
many countries in winter. This explanation may be correct for the epidemics
of flu that we observe in the present times
but a study of the influenza epidemics in
Iceland \cite{Weinberger2012} has shown that the timing of the epidemics has
changed over time. Prior to the early 1930 the influenza epidemics in Iceland
lacked a consistent seasonal pattern and sometimes included several peaks in
the summer months. This shows how the origin of the periodicity of epidemics
may be subtle. In 1957 Bartlett had already pointed out a clear influence of
the size of the cities on the periodicity of measles epidemics.

\medskip
The case of Covid-19 is particularly interesting because it is a {\em new
  disease} and we have data from its origin. For influenza the epidemics are
clearly affected by the immunity acquired by some individuals during previous
epidemics. This is why the analysis of the age distribution of the individuals
affected by an influenza epidemic
exhibits minima for specific age groups who acquired
immunity earlier \cite{Housworth1971}. This is not the case for the Covid-19
epidemic. Moreover the various mutants of the
Covid-19 virus have very different properties which bring specific clues for
the analysis of the data, allowing us to discriminate between different
possible explanations of the quasi-periodic outbursts of the epidemic.

In this study we use the data of two countries, France and Germany, which
have many similarities in their social and medical systems although they
differ in some of their approaches for the containment of the Covid-19
epidemic. Section \ref{sec:data} shows that these datasets
(Fig.~\ref{fig:covidcases}) have similar dynamics in 
the timing of the Covid-19 outbursts. We then consider two recent proposals
to explain the recurrence of outbursts. Section \ref{sec:memory} considers the
effect of the memory effects introduced by the finite duration of the acquired
immunity; Section \ref{sec:cluster} examines the role played by the saturation
of clusters within the population. We show how the data and the available
knowledge on the Covid-19 disease allow us to determine the most relevant
approach, and in  Sec.~\ref{sec:discussion} we discuss
our conclusions and their relation with earlier investigations
of other epidemics, and we emphasize some points worth keeping in mind.

\section{Dynamics of Covid-19 outbursts in France and Germany}
\label{sec:data}

\begin{figure}[ht]
  {\centering
  \includegraphics[width=8cm]{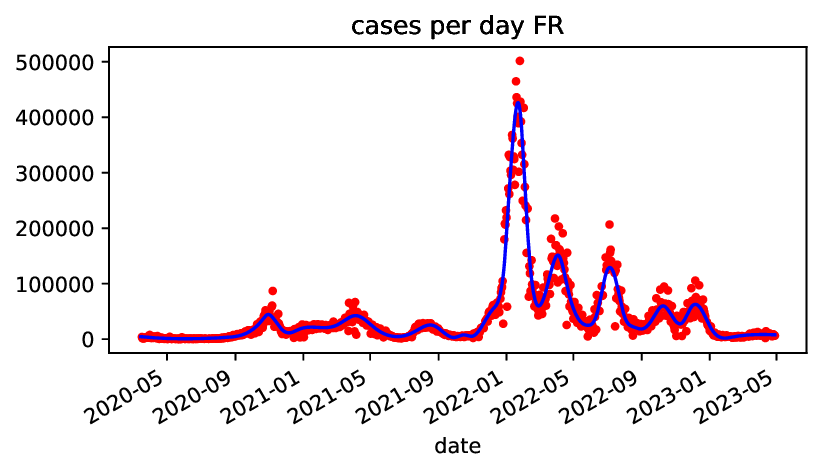} \\ 
  \includegraphics[width=8cm]{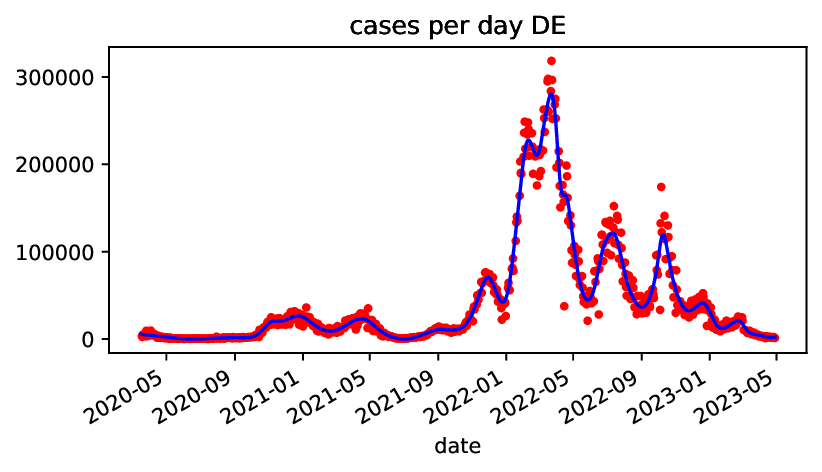} } 
  \caption{Number of new Covid-19 infections reported each day in France (top)
  and Germany (bottom) as a function of the date. The dots show the daily
  numbers (weekdays). As the data are very noisy the curves show the data
  smoothed by locally weighted smoothing followed by a cubic spline
  interpolation to get a set of data on a regular grid for further analysis. }
  \label{fig:covidcases}
\end{figure}

Figure \ref{fig:covidcases} shows the number of Covid-19 infections reported
each day in France and Germany from April 2020 to May 2023. These data deal
with large numbers, collected in the whole country, but they nevertheless
exhibit large fluctuations that are due to the reporting process. Therefore
we also plot curves that smooth out these fluctuations to display the general
features of the epidemic evolution more clearly.  During the reported period,
the epidemic never faded out but peaks corresponding to epidemic outbursts
emerge from a background corresponding to periods during which the
contamination was much weaker.

The general patterns are the same in both countries, in spite of some
differences in the rules imposed to contain the propagation of the epidemic
such as travel restrictions and regulations restricting shopping or mandating
face-masks for instance. Until the beginning of 2022 the peaks are smaller and
broader than in 2022 and 2023, and the interval between them is of the order
of 130--150 days. From the beginning of 2022, the peaks become much more
intense and their intervals decay to values of the order of 90--95 days,
and may even drop around 70 days. The
peaks cannot be associated to surges due to cold seasons since there are
maxima in April 2021, August 2021, April 2022, July 2022 in France and April
2021, early September 2021, April 2022, July 2022 in Germany. Therefore
Covid-19 outbursts are different from the seasonal outbursts observed nowadays
for influenza.

\medskip
To get a more quantitative and systematic view of the dynamic of the epidemics
in the two countries, we made a time-frequency analysis of the data by means
of the Wigner-Ville (WV) transform \cite{Flandrin1982, Flandrin1985, pytftb}
applied to the smoothed and interpolated data
plotted as continuous curves in Fig.~\ref{fig:covidcases}. This transform can
be viewed as an improved version of the short-term Fourier transform to detect
frequencies present in a signal in a finite time domain. It is appropriate for
non-stationary signals and reduces the effect of the crippling compromise
between the resolution in time and frequency
\cite{Flandrin1982,Flandrin1985}. Owing to its broad use in signal processing
software packages implementing a discrete-time version of the WV
transform \cite{pytftb} have been developed.

\begin{figure}[ht]
  {\centering
  \includegraphics[width=8cm]{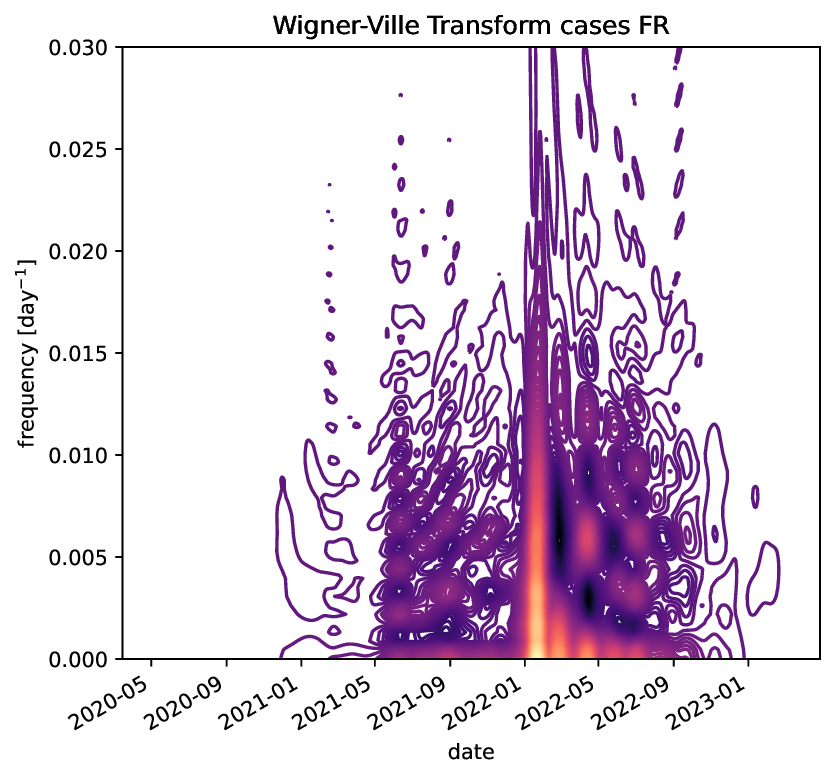} \\ 
  \includegraphics[width=8cm]{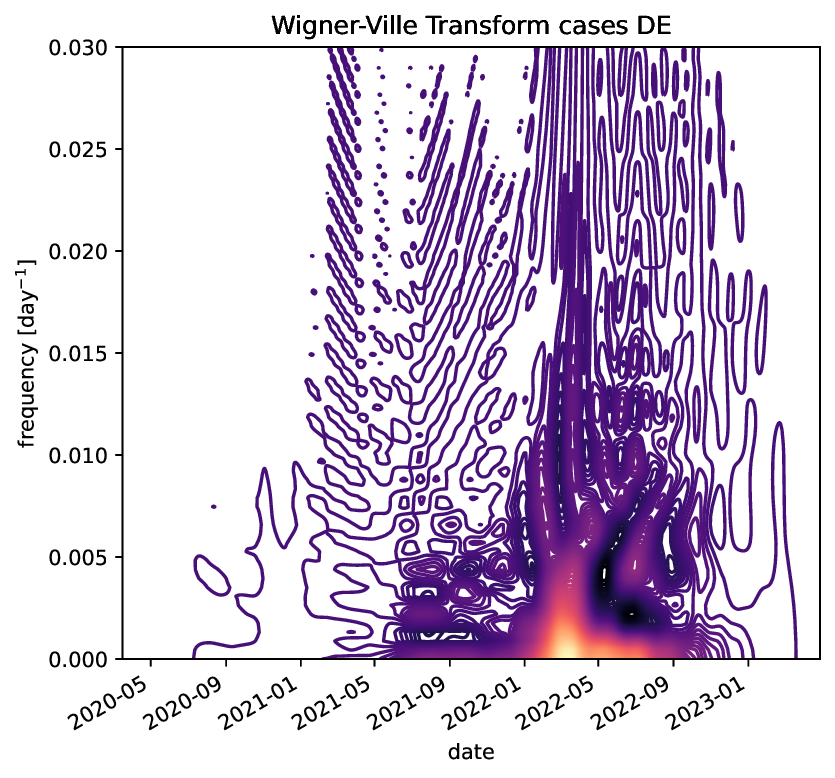}} 
  \caption{Time-frequency analysis of the smoothed and interpolated data
plotted as continuous curves in Fig.~\protect\ref{fig:covidcases}, using the
Wigner-Ville transform. The figure shows contour-plots of the
modulus of the Wigner-Ville
transform of the signal in France (top picture) and Germany (bottom) in the
date-frequency plane. The scale extends over the full range of the modulus of
the Wigner-Ville spectrum with 80 contour levels. The largest values,
observed around the beginning of year 2022, are highlighted by the
orange/yellow filling of the contours.}
  \label{fig:wvtransforms}
\end{figure}

The WV spectra of the smoothed curves showing the variation versus time of
the daily numbers of Covid-19 infections in France and Germany
(Fig.\ref{fig:covidcases}) are plotted in Fig.~\ref{fig:wvtransforms}. The
similarity of the WV time-frequency spectra in France and Germany is striking,
in spite of the differences in the shapes of the time-dependent curves of
Fig.\ref{fig:covidcases}. A simple glance at those WV spectra already tells a
lot on the Covid-19 infection, especially if we confront the spectra with the
spread of the virus variants in Europe, well represented (up to a few days) by
the knowledge for France. The first significant change, following the original
$\alpha$ and $\beta$ forms, was the appearance of the $\delta$ variant, which
had already caused a large epidemic in India. In France the $\delta$ variant
was at the origin of 34\% of the cases in June 2021 and 99\% in August
2021. The Omicron variant contributed to 40\% of the new infections
mid-December 2021 and more than 95\% in January 2022. The large extension of the
WV-spectra toward high frequencies, which appears clearly at the very
beginning of 2022 in France and Germany could have been guessed from the
curves of Fig.\ref{fig:covidcases} because it is due to the strong, sharp,
peaks that are observed at that time and were explained by the presence of
the Omicron variant. But the detection of the arrival of $\delta$ variant
would not be noticed on Fig.\ref{fig:covidcases}, while it appears on the WV
spectra by a significant rise in high-frequency contributions around June 2021
(particularly in France, but in Germany as well).

This shows that the WV-time-frequency analysis of the epidemic data may be a
useful tool to follow the development of an epidemic. This result is
particularly interesting to test epidemic models because it tells us that a
successful model should be able to show the kind of change in the period of
the recurrence of outbursts which is observed in the French and German
epidemics provided its parameters are modified in a manner quantitatively
consistent with the known properties of the virus variants. We show below that
not all models are able to withstand this test.

\section{Memory effects}
\label{sec:memory}

The basic epidemic model is the SIR model
introduced and studied in details by Kermack and McKendrick who proposed an
approximate solution for the profile of the epidemic \cite{Kermack1927}. In
this model, once an individual has recovered he acquires a permanent immunity,
so that the number of susceptible individuals is progressively exhausted and
the infection peak dies-out completely. The infection never restarts.

One possible source of periodic outbursts is that actually immunity does not
last forever. In the case of the Covid-19 disease, it disappears after a
finite-time interval. This introduces a memory effect in the dynamics of the
epidemic, which can lead to recurrences. An extension of the SIR model
to take into account this finite-time immunity was
recently introduced by Bestehorn et al. \cite{Bestehorn2022}. We implemented
this model to test its suitability to describe the dynamics of the Covid-19
epidemics in France and Germany. This section presents our results after a
brief introduction to the model.

\subsection{The extended SIR model with memory}

Let us denote by $S$, $I$ and $R$ the fraction of susceptible, infected and
recovered individuals in the total population so that $S + I + R = 1$. The
simple SIR model assumes that an infection occurs with rate $\beta$ when an
infectious person and a susceptible individual come into contact,
and that an infected
individual recovers (or more precisely stops being infectious)
after an average time $\gamma$. The dynamics of the model
versus time $t$ is, therefore,
described by the set of equations
\begin{equation}
  \label{eq:SIR1}
  \frac{dS}{dt} = - \beta \, I \, S ,  \quad
  \frac{dI}{dt} =  \beta \, I \, S - \gamma I , \quad
  \frac{dR}{dt} = \gamma I \; .
\end{equation}
It is convenient to introduce a dimensionless time $t' = t / \gamma$ and the
basic reproduction number $\rho = \beta / \gamma$, so that, in these units, the 
model depends on the single parameter $\rho$ 
\begin{equation}
  \label{eq:SIR2}
  \frac{dS}{dt'} = - \rho \, I \, S ,  \quad
  \frac{dI}{dt'} =  \rho \, I \, S - I , \quad
  \frac{dR}{dt'} =  I \; .
\end{equation}
We, henceforth, drop the prime for the time. It is understood that the unit
of time is the average time needed for an individual to recover.

\begin{figure}[ht]
  \centering
  \includegraphics[width=7cm]{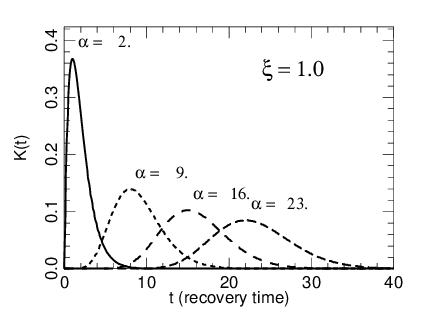} 

  \includegraphics[width=7cm]{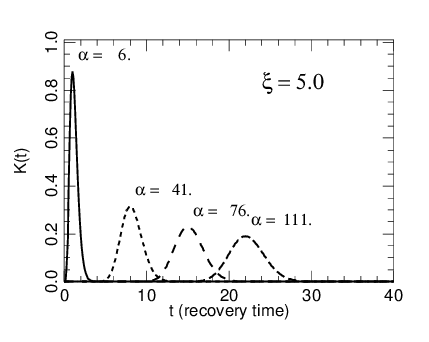} 
  \caption{The Erlang distribution for selected values of $\alpha$ and $\xi$.}
  \label{fig:erlang}
\end{figure}

\medskip
The model with memory assumes that an individual who recovered at time $t$
loses its immunity after a delay $\tau$ with probability $K(\tau)$ and becomes
susceptible again. The equations are, therefore, completed by extra terms to
describe the decay in the population of recovered individuals and the
corresponding growth in the population of susceptible individuals.
\begin{subequations}
  \label{eq:mmodel}
  \begin{align}
  \frac{dS(t)}{dt} &= - \rho \, I(t) \, S(t) +
  \int_{0}^{\infty} I(t - \tau) K(\tau) d\tau \label{eq:mmodela} \\
  \frac{dI(t)}{dt} &=  \rho \, I(t) \, S(t) - I(t) \label{eq:mmodelb} \\
  \frac{dR(t)}{dt} &=  I(t) -
  \int_{0}^{\infty} I(t - \tau) K(\tau) d\tau \label{eq:mmodelc}
  \end{align}
\end{subequations}
As suggested by Bestehorn et al. \cite{Bestehorn2022}, a convenient functional
form for $K(t)$ is the Erlang distribution
\begin{equation}
  \label{eq:erlang}
  K_{\alpha,\xi}(t) = \frac{\xi^{\alpha} t^{\alpha - 1}}{\Gamma(\alpha)}
  e^{-\xi t} , \quad \alpha > 0, \quad \xi > 0, \quad t \ge 0
\end{equation}
plotted in Fig~\ref{fig:erlang}. $\Gamma(\alpha)$ is the Euler $\Gamma$
function, $\alpha$ is a positive real number and $\xi$ is a scaling factor for
time, with
larger $\xi$ values making the distribution sharper. The maximum of the
distribution, which is the typical lifetime of the immunity is 
reached for
\begin{equation}
  \label{eq:tau0}
  t = \tau_0 = (\alpha - 1)/\xi \quad.
\end{equation}

\subsection{The Covid-19 infection described by the memory model}

The model described by Eqs.~\ref{eq:mmodel} does not have an
analytical solution,
but it can easily be investigated by numerical simulations. We assume that the
fraction of infected individuals is $I(t) = 0$ for $t < 0$
and $I(t=0) = I_0 = 2 \, 10^{-2}$,
$S(t=0) = 1 - I_0$ and $R(t=0) =
0$. As $I(t)$ vanishes for $t<0$, the integrals of Eqs.~\ref{eq:mmodel} have to
be calculated in the domain $0 \le \tau < t$. We use a fourth-order
Runge-Kutta algorithm to solve Eqs.~\ref{eq:mmodel}, with a
time step $dt = 0.01$
in units of the recovery time $\gamma$.

\begin{figure}[ht]
  \centering
  \includegraphics[width=8cm]{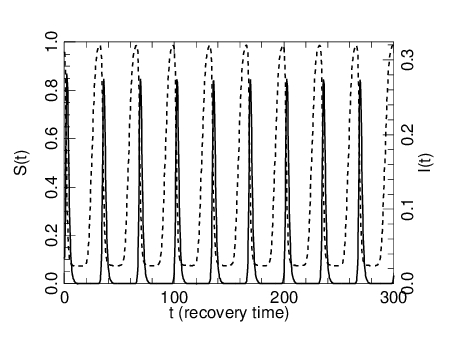} 
  \caption{Fraction of susceptible individuals (dashed line, left scale) and
    infected individuals (full line, right scale) versus time for the memory
    model with $\alpha = 111$, $\xi = 5.0$ and $\rho = 2.8$.}
  \label{fig:calc1024}
\end{figure}

\begin{figure}[ht]
  \centering
  \includegraphics[width=7cm]{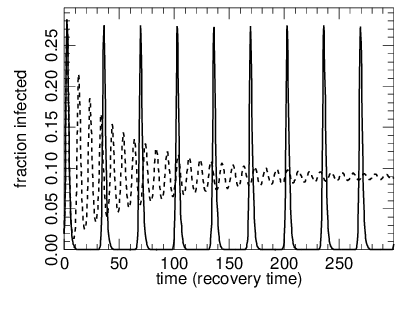} 
  \caption{Variation with time of the fraction of infected individuals for two
    parameter sets of the model: full line same as in Fig.~\ref{fig:calc1024}
    corresponding to $\tau_0 = 22$ (recovery time) and dashed line $\alpha =
    41$, $\xi = 5.0$, $\rho = 2.8$
    corresponding to $\tau_0 = 6$ (recovery time)}
  \label{fig:calc1024-1026}
\end{figure}

The model has two stationary solutions $0 \le \overline{S} \le 1$,
$\overline{I} = 0$, $\overline{R} = 1 - \overline{S}$ which corresponds to a
healthy population and $\overline{S} = 1/\rho$, $0 \le \overline{I} \le 1 -
\overline{S}$ which corresponds to a stable endemic situation. For a given
value of $\rho$, above a
bifurcation point that can be reached by increasing $\tau_0$,
the stable endemic point becomes unstable and this leads to a
series of periodic outbursts as shown in Fig.~\ref{fig:calc1024}
When $\tau_0$ decreases, the period between the outbursts gets shorter, and,
below the bifurcation point the outbursts decay with time and the model
evolves toward the endemic stable point as shown in
Fig.~\ref{fig:calc1024-1026}.
Although we did not try to fit observed data because the model is too crude to
pretend to be realistic, the parameters used to generate
Fig.~\ref{fig:calc1024} have been chosen to give plausible numbers for the
first stage of
the epidemic in France and Germany (with $\alpha$ and $\beta$
virus variants). The recovery time $\gamma$, which is the time during which an
infected individual can contaminate susceptible individuals can be estimated
to be 5 days, so that the mean immunity lifetime $\tau_0 = 22$, measured in
terms of the recovery time is of the order of 3.7 months with a standard
deviation of about 20 days, as shown in Fig.~\ref{fig:erlang} by plotting the
probability distribution function for the lifetime of the immunity following
the disease. The period between the outbursts on Fig.~\ref{fig:calc1024} is
$33.37\,\times\,$the recovery time, i.e.\ 166.8 days, which is the order of
magnitude of the time separating outbursts at the very beginning of the
epidemic. Choosing parameters leading to a larger value for
$\tau_0$, i.e.\ a longer immunity lifetime that is not
implausible, would bring the period between the
outbursts well above the observed values, which already casts a doubt on the
validity of this model for the Covid-19 epidemics.

Evaluation of the model parameters that depend on the virus variant is
not straightforward 
because many of the studies on infectivity or duration of the immunity are
derived from biological parameters such as the viral load or the persistence
of antibodies in the serum, rather than the influence of the variants on the
epidemic itself. There are, nevertheless, some studies which provide some useful
insight. A paper by Liu and Rockl\"ov \cite{Liu2021} compares the reproductive
number for the $\delta$ variant, evaluated on average to be in the range $3.2
\le \rho \le 8$ with a mean of $5.08$, to the value for the original strain
$\rho = 2.79$. This is why we selected $\rho = 2.8$ for the calculation
presented in Fig.~\ref{fig:calc1024} meant to describe the first stage of the
epidemic. The lifetime of the immunity has been often evaluated for the
vaccine-acquired immunity, which could be different from the immunity acquired
from the disease, but may be expected to be similar. The main point for our
study is that for Omicron the immunity is much shorter than for the variants
present in the first stage of the epidemic \cite{Andrews2022}. It decreases
very significantly after 10 weeks, but many cases of multiple fast
re-infections by Omicron have been observed. On average, taking into account
the cases of individuals who hardly appeared to have any immunity we can
estimate that the immunity provided by an Omicron infection does not exceed a
month.

\begin{figure}[ht]
  \centering
  \includegraphics[width=7.5cm]{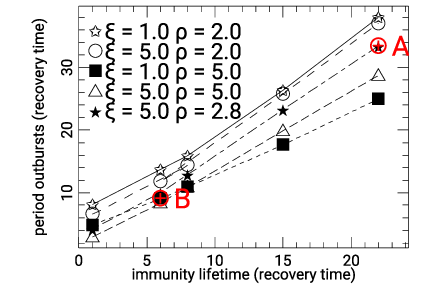} 
  \caption{Results for the memory model for a variety of parameters. The
    values of $\xi=1$ and $\xi = 5$, illustrate the effect of the
    sharpness of the distribution of the lifetime of the immunity. The value
    $\xi = 5$, which leads to a standard deviation of about 3 weeks for the
    lifetime of the immunity is the most relevant for the Covid-19
    epidemic. The reproduction number $\rho$ determines the average number of
    new infections generated by an infectious person. It was estimated to be
    $\rho \approx 2.8$ for the ancestral strain of the virus and to rise to
   $\rho \approx 5$ for the $\delta$ variant \cite{Liu2021}, and is probably
   of the same order for Omicron. The lines connect data with the same $\xi$
   and $\rho$ values, but a varying immunity lifetime $\tau_0$ obtained by
   varying $\alpha$ according to Eq.~(\ref{eq:tau0}). The points marked {\bf
     A} and {\bf B} show the parameters chosen as appropriate to
   simulate first and second stages of the Covid-19 epidemic.}
  \label{fig:memoryres}
\end{figure}

Figure \ref{fig:memoryres}, which shows how the period of the outbursts
varies with the lifetime of the immunity in the case of the memory model can
be used to analyze the validity of this model for the Covid-19 epidemic in
France and Germany. First, one should notice that the sharpness of the
distribution of immunity lifetimes is not a crucial feature. For $\rho = 2$,
the periodicities of the outbursts for $\xi = 1$
(open stars, broad distribution) and
$\xi = 5$ (open circles, sharper, more realistic distribution) are almost the
same and only differ slightly for the shortest immunity lifetimes. For $\rho =
5$, the results (closed squares and open triangles) differ a bit more but
nevertheless not very much. To describe the two stages of the epidemic
observed in France and Germany, the relevant points are points {\bf A} ($\xi =
5$, $\rho = 2.8$, $\tau_0 = 22\,\times\,$ the recovery time, i.e. $\approx 3.7$
months corresponding to $\alpha$ or $\beta$ virus variants) that gives
a period of $P_1 = 33.37\,\times\,$the recovery time,
i.e. $P_1 \approx 167$ days, and
{\bf B} ($\xi =
5$, $\rho = 5$, $\tau_0 = 6\,\times\,$the recovery time,
i.e. $\approx 30$ days,
corresponding to Omicron) that gives a period  $P_2 = 9.07\,\times\,$the
recovery
time, i.e. $P_2 \approx 45$ days. The ratio $P_1 / P_2 \approx 3.7$ is much
higher than the ratio of less than 2 that is observed when Omicron
replaces earlier variants in France and Germany. Moreover, as observations have
shown that the Omicron infection offers a very weak protection against a
resurgence of the disease for an individual, our estimate of $30$ days for the
lifetime of the Omicron immunity is probably an overestimate,
while the immunity lifetime for the original virus strain and $\alpha$ and
$\beta$ variants that we selected to be $3.7$ months may be
underestimated. Therefore the ratio $P_1/P_2$ 
provided by the memory model could probably be even higher. In fact
Fig.~\ref{fig:memoryres} shows that, for a given $\rho$ the period of the
outbursts predicted by the model is almost proportional to the immunity
lifetime, which is the the basic idea behind the memory model. As the
observation of the epidemics has shown that the immunity provided by the
Omicron infection is very much weaker than the immunity provided by the initial
strains of the virus, it could have been predicted that the memory model
should lead to a very large change in the dynamics of the epidemics when the
Omicron variant started to dominate. The observations
do not show such a large a qualitative transition in the periods of the
outbursts, {\em suggesting that the memory effect due to the lifetime of the
immunity is not at the origin of the periodic outbursts for the Covid-19
epidemics} although one can find some model parameters that appear to fit a
few successive outburst of a Covid epidemic \cite{Bestehorn2022}.
Testing the evolution due to the
virus variants provides a harder test that the model does not seem to be able
to pass with parameters that stay in a realistic range.

\section{Cluster saturation}
\label{sec:cluster}

Another approach was recently put forward by Gostiaux et al.\
\cite{Gostiaux2023} to explain periodic epidemic outbursts with an extension
of the SIR model. The idea is that the propagation of the disease does not
affect the population globally but instead that it is controlled by the events
in local clusters that interact weakly with the global population that
plays the role of a reservoir. In this approach a SIR-like model is written at
the level of a cluster. Some of the individuals who recover within the cluster
can be transferred to a group of recovered individuals in the general
population while some members of the general population, who had not yet been
affected by the disease can join the group of susceptible individuals of the
cluster, which, therefore, tends to grow.
This last hypothesis is actually the crucial point, which is at the
origin of periodic outbursts.

\medskip
In fact the idea that the growth of the group of the susceptible individuals
was behind periodic outbursts of an epidemic was already introduced in 1929 by
Soper who studied ``{\em The interpretation of periodicity in disease
  prevalence}'' \cite{Soper1929}. He was investigating the measles epidemics,
and wrote: ``that the accumulation
of susceptibles -- since more than 90\% of all
children born in Western Europe and surviving infancy pass through an attack
of measles -- is an important factor of the oscillations or periods of the
epidemics has been adopted by the great majority of epidemiologists''. Then he
presented an analysis showing that the growth of the number of susceptible
individuals is enough to give rise to periodic oscillations. In his paper he
considered several models because, for measles, the infectious period is short
and follows a larger period of incubation. But, in the simplest case where the
infectious period can be assumed to extend during the whole illness, Soper came
to a model that is analogous to the SIR model completed by a single term that
generates a continuous growth of the population of susceptible
individuals. With our notations, Soper's model is described by the equations
\begin{subequations}
  \label{eq:cmodel}
  \begin{align}
  \frac{dS(t)}{dt} &= - \rho \, I(t) \, S(t) + \epsilon \label{eq:cmodela} \\
  \frac{dI(t)}{dt} &=  \rho \, I(t) \, S(t) - I(t) \label{eq:cmodelb}
  \end{align}
\end{subequations}
when the time is measured in units of the recovery time.
For measles the extra term $\epsilon$ added to the first equation of the SIR
model was simply explained by the birth rate of children in a given area
(typically an English city) affected by periodic outbursts of measles.
In the context of the Covid-19 epidemic, we, henceforth, call
this model the ``cluster model'' because it can be viewed as a simplified
version of the model of Gostiaux et al. \cite{Gostiaux2023}. It should be
understood as the model of a sub-population (the ``cluster'') that is in
contact with a broader reservoir of population. In their approach Gostiaux et
al.\ described the interaction as a diffusion process in two dimensions, with
a cluster population viewed as proportional to the area of the domain covered
by the sub-population and a rate of change that is determined by the
displacement of the perimeter of the domain. In two-dimensional diffusion
a segment of the perimeter moves as $\sqrt{t}$ so that the area of the domain
grows proportionally to $t$.
This growth corresponds to $dS/dt = \epsilon$, where $\epsilon$
is simply the proportionality constant, hence the term added to the SIR
model in Eq.~(\ref{eq:cmodela}).

\medskip
In this section we show that this highly simple model, with very few
parameters, is sufficient to explain important observations on the Covid-19
epidemics. 
\begin{figure}[ht]
  \centering
  \includegraphics[width=8cm]{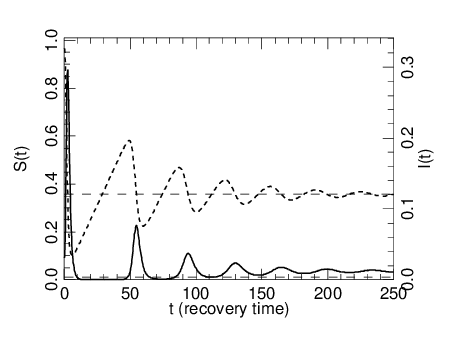} 
  \caption{Fraction of susceptible individuals (dashed line, left scale) and
    infected individuals (full line, right scale) versus time for the cluster
    model with $\rho = 2.8$ and $\epsilon = 0.012$. The horizontal long-dashed
  thin lines show the values corresponding to the fixed point $\overline{S} =
  1/\rho$ and $\overline{I} = \epsilon$.}
  \label{fig:cmodel-1010}
\end{figure}
Figure \ref{fig:cmodel-1010} shows the variation versus time of the fraction
of susceptible (dashed lines) and infected (full line) individuals computed by
the cluster model with $\rho = 2.8$ and $\epsilon = 0.012$. As for the memory
model the simulation was carried out by a numerical integration of
Eqs~(\ref{eq:cmodel}) with a fourth-order Runge-Kutta algorithm with a time
step $dt = 0.01$ in units of the recovery time. The initial condition assumes a
fraction of infected individuals $I(t=0) = I_0 = 2\, 10^{-2}$, $S(t=0) = 1 -
I_0$ and $R(t=0) = 0$. The model gives a sequence of successive outbursts
that decay as time grows. The pattern is qualitatively similar to the series
of outbursts observed in France and Germany in 2022 and 2023
(Figs.~\ref{fig:covidcases}). In the long term the calculation reaches an
endemic state with a small fraction of infected individuals. Each outburst is
preceded by a quasi-linear growth of the fraction of susceptible individuals,
which drops sharply when an outburst occurs, before growing again.

\medskip
In spite of the apparent simplicity of the model, we could not derive an
analytical solution to the set of nonlinear equations
(\ref{eq:cmodel}). However some of the main features of the numerical solution
can be deduced from these equations.

First, the model has a fixed point $\overline{I}$ given by
\begin{equation}
  \label{eq:fixedcm}
  \frac{d \overline{S}}{dt} = 0 = - \rho \overline{S} \; \overline{I} + \epsilon
  \; , \quad \frac{d \overline{I}}{dt} = 0 = \rho \overline{S} \; \overline{I} -
  \overline{I}
\end{equation}
corresponding to $\overline{I} = \epsilon$ and $\overline{S} = 1/\rho$ which
are shown by long-dashed horizontal lines on Fig.~\ref{fig:cmodel-1010}.

Due to the analogy between the SIR model and the cluster model, the dynamics
of an outburst can be approximately described by the solution obtained by
Kermack and McKendrick \cite{Kermack1927} for the SIR model in the limit
$\rho R \ll 1$. Using the third equation in Eqs.~(\ref{eq:SIR1}), the first of
these equations becomes $dS/dt = - \rho S \; dR/dt$.
Using the initial condition
$R(t=0) = 0$, $S(t=0) = S_0$ $I_0 = 1 - S_0$, one obtains
\begin{equation}
  \label{eq:SR}
  S = S_0 e^{-\rho R} \,
\end{equation}
and then, in the case $\epsilon = 0$ Kermack and McKendrick
could use the condition
$I + S + R = 1$ so that the third equation of Eq.~\ref{eq:SIR1} finally gives
\begin{equation}
  \label{eq:eqR}
  \frac{dR}{dt} = 1 - S_0 e^{-\rho R} - R \; .
\end{equation}
This nonlinear differential equation for $R(t)$ cannot be solved exactly, but,
in the limit $\rho R \ll 1$, expanding the exponential up to second order
gives
\begin{equation}
  \label{eq:eqR2}
  \frac{dR}{dt} = 1 - S_0 + (\rho S_0 - 1) R - \frac{1}{2} \rho^2 S_0 R^2 \; .
\end{equation}
A second order expansion was required because $\rho R$ has to be
compared to $1 - S_0 = I_0$ which is itself small. 
Equation (\ref{eq:eqR2}) has the solution \cite{Kermack1927}
\begin{equation}
  \label{eq:solR}
  R(t) = \frac{1}{\rho^2 S_0} \left[ \rho S_0 - 1 + \nu
    \tanh \left( \frac{\nu t}{2} - \phi \right) \right] \; ,
\end{equation}
with $\nu^2 = (\rho S_0 - 1)^2 + 2 S_0 I_0 \rho^2$ and $\phi = \tanh^{-1}
[(\rho S_0 - 1)/\nu]$. This gives the shape of an outburst as
\begin{equation}
  \label{eq:solI}
  \frac{dR(t)}{dt} = I(t) = \frac{\nu^2}{2 \rho^2 S_0} \mathrm{sech}^2 \left(
    \frac{\nu t}{2} - \phi \right) \; .
\end{equation}
\begin{figure}[ht]
  \centering
  \includegraphics[width=7cm]{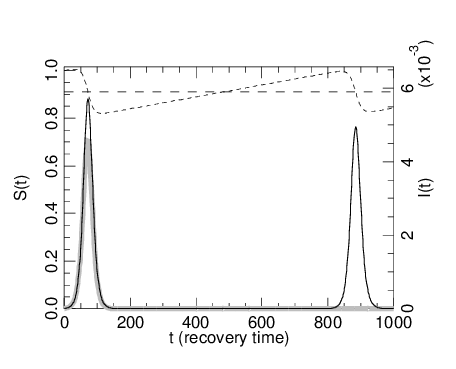} 
  \caption{Fraction of susceptible individuals (short-dashed line, left scale)
    and
    infected individuals (full line, right scale) versus time for the cluster
    model with $\rho = 1.1$ and $\epsilon = 2.5\,10^{-4}$. The horizontal
    long-dashed line shows the value of the fixed point for the fraction of
    susceptible individuals $\overline{S} = 1/\rho$. The thick gray full line
  shows the theoretical solution for the SIR model in the limit $\rho R \ll
  1$.} 
  \label{fig:theosir}
\end{figure}
Figure \ref{fig:theosir} shows that the theoretical shape of the outbursts
derived for the SIR model in the limit $\rho R \ll 1$ provides a fairly good
description of the numerical result obtained for the cluster model for a small
value of $\epsilon = 2.5\, 10^{-4}$. This shows that the analytical results of 
Kermack and McKendrick \cite{Kermack1927} for the SIR model can be used to get
some insight into the properties of the cluster model. The important point of
their analytical investigation, besides the shape of the outbursts, is that it
shows that the epidemic outburst is triggered by a threshold in the fraction
of susceptible individuals, which, for the SIR model is equal to
$\widetilde{S} = 1/\rho$. The simulation results (Fig.~\ref{fig:theosir})
indicate that there is still a threshold for $\epsilon \not= 0$.
However its value
is moved up to $\widetilde{S'} = 1/\rho + \widetilde{\Delta S}$.
This is the existence of this threshold that
explains the periodicity of the outbursts for $\epsilon \not= 0$. The
term $\epsilon$ in $dS/dt$ causes a growth of $S(t)$ until it reaches
the threshold and triggers an epidemic outburst. The outburst induces a quick
drop in $S(t)$, which brings it below the threshold, down to a value
approximately equal to $1/\rho - \widetilde{\Delta S}$.
After the outburst the growth of $S(t)$ restarts until the threshold is again
reached, causing the next outburst, and so on. For larger values of
$\epsilon$, the analytical result of \onlinecite{Kermack1927} is no longer
quantitatively valid. Figure \ref{fig:cmodel-1010} shows that the threshold 
$\widetilde{S'}$ tends to decrease from one outburst to the next, presumably
because it depends on the value of $I(t)$ at the end of the previous outburst.

\begin{figure}[ht]
  \centering
  \includegraphics[width=7cm]{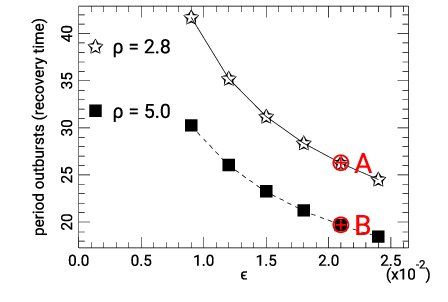} 
  \caption{Results for the cluster model for a variety of parameters. The two
    values of $\rho$ correspond to the estimated values of the average number
    of new infections generated by an infectious person in the first stage
    ($\rho = 2.8$) and second stage of the Covid-19 epidemic ($\rho=5$)
    \cite{Liu2021}. The figure shows how the period of the outbursts depend on
    $\epsilon$ for those two values of $\rho$. Points {\bf A} and {\bf B} mark
  the parameters appropriate for the first and second stage of the epidemic.}
  \label{fig:cmodelres}
\end{figure}
Figure \ref{fig:cmodelres} shows how the period of the outbursts, measured in
units of the recovery time, depends on $\epsilon$ for the two values
$\rho=2.8$ and $\rho = 5$ corresponding to first and second stages of the
Covid-19 epidemic (see Sec.~\ref{sec:memory} and \cite{Liu2021}).
 For a fixed
 $\widetilde{S'}$ threshold, the period should decrease linearly with the
 increase of $\epsilon$. A weak deviation from linearity, observed on
 Fig.~\ref{fig:cmodelres}, occurs because $\widetilde{S'}$ depends on
 $\epsilon$.

 The parameter $\epsilon$, which determines the growth of the number of
 susceptible individuals in a cluster, is a measure of the interactions
 between people in the society. We can assume that, on average, it did not
 vary significantly during the full epidemic, and moreover, as social
 behaviors are similar in France and Germany, we can
 select the same value of $\epsilon$ for the two countries. Selecting
 $\epsilon = 0.021$ corresponds to the points {\bf A} (first stage of the
 epidemic, $\rho = 2.8$) and {\bf B} (second stage, $\rho = 5$) marked on
 Fig.~\ref{fig:cmodelres}. With this value of $\epsilon$, the periods of
 the outbursts determined by the cluster model are $P_1 = 26.23\,\times\,$the 
 recovery time, i.e.\ $P_1 \approx 131$ days for the first stage of the
 epidemic and $P_2 = 19.67\,\times\,$the recovery
 time, i.e.\ $P_2 \approx 98$ days
 for the second stage. These two numbers are in good agreement with the
 numbers observed in France and Germany, and, contrary to what we had noticed
 for the memory model, $P_1/P_2 \approx 1.34$ is consistent with the
 observations. It should be stressed that, owing to the simplicity of the
 cluster model that only has two parameters $\rho$ and $\epsilon$, there
 is little room to fiddle with parameters to reach an agreement between
 the theory
 and experiment. The values of $\rho$ in both stages of the epidemic result
 from medical observations \cite{Liu2021}. They are certainly not perfectly
 accurate, but they are in line with the observations in various
 countries. The only free parameter of the model is $\epsilon$. Changing
 $\epsilon$ modifies $P_1$ and $P_2$ as shown in Fig.~\ref{fig:cmodelres}, but
 the ratio $P_1/P_2$ stays almost constant. Therefore there is actually no
 freedom for data fitting. Once $\epsilon$ is chosen to get a correct value
 of $P_1$, then $P_2$ is determined. As the model, with realistic values for
 $\rho$, gives a ratio $P_1/P_2$ which agrees with the observations, it
 strongly suggests that the mechanism behind the periodic outbursts of the
 Covid-19 epidemics is the growth of the clusters of affected individuals, due
 to social contacts, rather than the finite duration of the immunity to
 the virus provided by the disease (or by vaccines) as assumed for the memory
 model. 

\section{Discussion}
\label{sec:discussion}

Although a large number of investigations have been devoted to the modeling of
epidemics, some basic questions remain open. Accurate modeling of specific
events requires a very accurate description, involving many intricate
phenomena, and, therefore, many parameters that have to be fitted. Such models
may help in managing epidemic containment and disease treatment. However, the
difficulty is sometimes to identify the real cause behind the observations.

This is the case for the periodic outbursts of epidemics. Seasonal factors
have been invoked, but they do not apply to the dynamics of Covid-19 epidemics
observed in Europe (Sec.~\ref{sec:data}). Moreover historic studies of the
influenza epidemics, which are now seasonal through the world, have shown that
earlier epidemics in Iceland did not have such a seasonal character
\cite{Weinberger2012} indicating that the phenomena may be subtle. Two main
ideas to explain the periodic outbursts of the Covid-19 epidemics have
recently been explored. One relies on the finite duration of the acquired
immunity \cite{Bestehorn2022}, while the second considers the role of
epidemic clusters \cite{Gostiaux2023}. Determining which one is the
appropriate mechanism is not easy because it is always possible to find a
parameter set that can match one observed periodicity.

However we have shown that the dynamics of the Covid-19 epidemics in France
and Germany is more complex than a single periodicity. In any case,
due to the many
phenomena that enter into play we should not expect a single, well-defined
period for the outbursts. However the data in both countries show a
clear pattern of periodicity change, correlated to the appearance of new virus
variants (Sec.~\ref{sec:data}). In a first stage of the epidemics, with the
original virus strain or $\alpha$ or $\beta$ variants, the periodicity of the
outbursts was about $1.3$--$1.5$ times longer than during the second stage
with the $\delta$ and Omicron variants. This observation is valid for two
independent data sets for France and Germany, and the transition between the
two stages is clearly visible in a time-frequency analysis of the infection
data. This provides a clear signal that might discriminate between the
possible models.

We used two models that have a small number of parameters. Moreover
these parameters, such as the duration of
the acquired immunity, or the number of individuals contaminated by somebody
who had been infected, can be rather accurately estimated from the medical data.
Our results show that the duration of the acquired
immunity is unlikely to explain the observations because the properties of the
Omicron variant that dominates the second stage of the epidemic are so much
different from those of the original strain or of the $\alpha$ and $\beta$
variants that they would lead to a change in the periodicity which is much
larger than the factor $1.3$--$1.5$ that is observed
(Sec.~\ref{sec:memory}). Instead a model based on the limited size of epidemic
clusters, which may grow by exchange with the reservoir provided by the general
population, appears to give a rather good agreement with the observations,
suggesting that the ideas behind this model are the source of the Covid-19
periodic outbursts (Sec.~\ref{sec:cluster}). As this model is very simple,
with only two parameters, one being determined by medical data and the second
controlling the periods in the two stages but not their ratio, there is very
little room to fiddle with parameters to reach an agreement with the
observations.

\bigskip
The results leave us with several points to keep in mind.

\medskip
$\bullet$ Do not forget old studies! The SIR model for epidemics, which is
behind many recent investigations,  has been 
thoroughly studied in 1927 by Kermack and McKendrick \cite{Kermack1927}.
And the ``cluster model''  had already been introduced by Soper in 1929 in a
different context \cite{Soper1929} when he studied measles epidemics in
England. The study of Soper, who showed how the periodicity of the outbursts of
measles epidemics could be related to the growth of the number of susceptible
individuals, appears to be of broader validity. The notion of
``cluster'' was also considered more than half a century ago in the case of
measles by Bartlett \cite{Bartlett1957}. He showed how the periodicity of
measles epidemics was varying according to the community size by studying the
case of many English cities. For his analysis, Bartlett had introduced a cell
model, with diffusion between cells, so that the essential features of the
cluster model of Gostiaux et al. \cite{Gostiaux2023} were contained in his
approach. His work shows that, while the epidemics tend to fade away in small
communities, the disease tends to stay in an endemic state when a community
gets large enough, as it does in the simple cluster model that we
investigated. Considering the large degree of exchanges that take place
now-days in the world, the lesson of history is not optimistic concerning
eradication of the Covid-19 disease. Fortunately modern vaccination could
alter the course of the events!

\medskip
$\bullet$ Small is beautiful. It is always tempting to set up complex models
because they can be realistic. The models that we explore, and particularly
Soper's model used as a simplified cluster model, would not be able to
reproduce the curves that we showed in Sec.~\ref{sec:data}, but if we focus on
a specific feature of the data, which is generic and not trivial to reproduce
while taking realistic constraints into account, the very small number of model
parameters is an important strength because it does not allow to play with
the parameters to describe the observations. This is a reminder of the famous
sentence attributed to J. von Neumann by E. Fermi: ``With four parameters I
can fit an elephant, and with five I can make him wiggle his
trunk.''\cite{Dyson2004}.

\medskip
$\bullet$ Attributing the Covid-19 outbursts to the parameter $\epsilon$ of
the cluster models may be tested by other studies. This parameter actually
measures the degree of contacts within the society. Therefore it must actually
vary depending on some specific situations, such as vacancies causing more
travel or celebrations that tend to bring families together like Christmas. A
statistical analysis of the Covid-19 data to try to correlate them to such
events could help support the ``cluster model''. Such a study must, however, be
made with care. There are examples showing that people gathering can lead to
Covid-19 outbursts, such a religious gathering in France
and of the ``Rosenmontag'' period in
Germany in the early stage of the Covid-19 epidemic. A correlation of a single
outburst with a gathering is a trivial effect and only sufficient statistics on
the fluctuations of the intervals between outbursts could be meaningful.
Moreover small fluctuations of $\epsilon$ due to external conditions extending
to a full 
country (such as the date of the beginning of the virus spreading, or the dates
of the school vacations) could contribute to synchronize the evolutions in
different clusters. Otherwise, if the clusters were small at the scale of a
country and independent from each other, various local outbursts would
average out and no clear periodicity would be seen.

$\bullet$ Finally one can point out that, although the memory model and the
cluster model lead to different conclusions about the origin of the
periodicity in Covid-19 outbursts they actually share the same fundamental
feature: outbursts appear when the concentration of susceptible individuals
has sufficiently grown. In the memory model the growth is due to the loss of
immunity of previously immunized individuals, while in the cluster model it
occurs because new individuals come into contact with the cluster. Nevertheless
behind both models one finds the idea first introduced in a model by Soper
\cite{Soper1929}.

\begin{acknowledgments}
I would like to acknowledge Dr.\ med.\ Franz-Geert Hagmann (Karlsruhe) for
helpful suggestions and references.
\end{acknowledgments}



\nocite{*}
\bibliography{covidrefs}

\begin{thebibliography}{15}%
\makeatletter
\providecommand \@ifxundefined [1]{%
 \@ifx{#1\undefined}
}%
\providecommand \@ifnum [1]{%
 \ifnum #1\expandafter \@firstoftwo
 \else \expandafter \@secondoftwo
 \fi
}%
\providecommand \@ifx [1]{%
 \ifx #1\expandafter \@firstoftwo
 \else \expandafter \@secondoftwo
 \fi
}%
\providecommand \natexlab [1]{#1}%
\providecommand \enquote  [1]{``#1''}%
\providecommand \bibnamefont  [1]{#1}%
\providecommand \bibfnamefont [1]{#1}%
\providecommand \citenamefont [1]{#1}%
\providecommand \href@noop [0]{\@secondoftwo}%
\providecommand \href [0]{\begingroup \@sanitize@url \@href}%
\providecommand \@href[1]{\@@startlink{#1}\@@href}%
\providecommand \@@href[1]{\endgroup#1\@@endlink}%
\providecommand \@sanitize@url [0]{\catcode `\\12\catcode `\$12\catcode
  `\&12\catcode `\#12\catcode `\^12\catcode `\_12\catcode `\%12\relax}%
\providecommand \@@startlink[1]{}%
\providecommand \@@endlink[0]{}%
\providecommand \url  [0]{\begingroup\@sanitize@url \@url }%
\providecommand \@url [1]{\endgroup\@href {#1}{\urlprefix }}%
\providecommand \urlprefix  [0]{URL }%
\providecommand \Eprint [0]{\href }%
\providecommand \doibase [0]{http://dx.doi.org/}%
\providecommand \selectlanguage [0]{\@gobble}%
\providecommand \bibinfo  [0]{\@secondoftwo}%
\providecommand \bibfield  [0]{\@secondoftwo}%
\providecommand \translation [1]{[#1]}%
\providecommand \BibitemOpen [0]{}%
\providecommand \bibitemStop [0]{}%
\providecommand \bibitemNoStop [0]{.\EOS\space}%
\providecommand \EOS [0]{\spacefactor3000\relax}%
\providecommand \BibitemShut  [1]{\csname bibitem#1\endcsname}%
\let\auto@bib@innerbib\@empty
\bibitem [{\citenamefont {Jones}\ and\ \citenamefont
  {Strigul}(2021)}]{Jones2020}%
  \BibitemOpen
  \bibfield  {author} {\bibinfo {author} {\bibfnamefont {L.~M.}\ \bibnamefont
  {Jones}}\ and\ \bibinfo {author} {\bibfnamefont {N.}~\bibnamefont
  {Strigul}},\ }\bibfield  {title} {\enquote {\bibinfo {title} {Is spread of
  covid-19 a chaotic epidemic?}}\ }\href@noop {} {\bibfield  {journal}
  {\bibinfo  {journal} {Chaos, Solitons and Fractals}\ }\textbf {\bibinfo
  {volume} {142}},\ \bibinfo {pages} {110376} (\bibinfo {year}
  {2021})}\BibitemShut {NoStop}%
\bibitem [{\citenamefont {Campbell}\ and\ \citenamefont
  {Rose}(1983)}]{Campbell1983}%
  \BibitemOpen
  \bibfield  {author} {\bibinfo {author} {\bibfnamefont {D.}~\bibnamefont
  {Campbell}}\ and\ \bibinfo {author} {\bibfnamefont {H.}~\bibnamefont
  {Rose}},\ }\bibfield  {title} {\enquote {\bibinfo {title} {Proceedings of the
  international conference on order in chaos, held at the
  {Center-for-Nonlinear-Studies}, {Los Alamos}, {New Mexico}, {USA}, 24-28 may
  1982. preface},}\ }\href@noop {} {\bibfield  {journal} {\bibinfo  {journal}
  {Physica D}\ }\textbf {\bibinfo {volume} {7}},\ \bibinfo {pages} {R7--R8}
  (\bibinfo {year} {1983})}\BibitemShut {NoStop}%
\bibitem [{\citenamefont {Kermack}\ and\ \citenamefont
  {McKendrick}(1927)}]{Kermack1927}%
  \BibitemOpen
  \bibfield  {author} {\bibinfo {author} {\bibfnamefont {W.~O.}\ \bibnamefont
  {Kermack}}\ and\ \bibinfo {author} {\bibfnamefont {A.~G.}\ \bibnamefont
  {McKendrick}},\ }\bibfield  {title} {\enquote {\bibinfo {title} {A
  contribution to the mathematical theory of epidemics},}\ }\href@noop {}
  {\bibfield  {journal} {\bibinfo  {journal} {Proc. R. Soc. London, Ser. A}\
  }\textbf {\bibinfo {volume} {115}},\ \bibinfo {pages} {700--721} (\bibinfo
  {year} {1927})}\BibitemShut {NoStop}%
\bibitem [{\citenamefont {Soper}(1929)}]{Soper1929}%
  \BibitemOpen
  \bibfield  {author} {\bibinfo {author} {\bibfnamefont {H.~E.}\ \bibnamefont
  {Soper}},\ }\bibfield  {title} {\enquote {\bibinfo {title} {The
  interpretation of periodicity in disease prevalence},}\ }\href@noop {}
  {\bibfield  {journal} {\bibinfo  {journal} {J. Royal Statistical Society}\
  }\textbf {\bibinfo {volume} {92}},\ \bibinfo {pages} {34--61} (\bibinfo
  {year} {1929})}\BibitemShut {NoStop}%
\bibitem [{\citenamefont {Bartlett}(1957)}]{Bartlett1957}%
  \BibitemOpen
  \bibfield  {author} {\bibinfo {author} {\bibfnamefont {M.~S.}\ \bibnamefont
  {Bartlett}},\ }\bibfield  {title} {\enquote {\bibinfo {title} {Measles
  periodicity and community size},}\ }\href@noop {} {\bibfield  {journal}
  {\bibinfo  {journal} {J. Royal Statistical Society Series A}\ }\textbf
  {\bibinfo {volume} {120}},\ \bibinfo {pages} {48--70} (\bibinfo {year}
  {1957})}\BibitemShut {NoStop}%
\bibitem [{\citenamefont {Weinberger}\ \emph {et~al.}(2012)\citenamefont
  {Weinberger}, \citenamefont {Krause}, \citenamefont {M{\o}lbak},
  \citenamefont {Cliff}, \citenamefont {Briem}, \citenamefont {Viboud},\ and\
  \citenamefont {Gottfredsson}}]{Weinberger2012}%
  \BibitemOpen
  \bibfield  {author} {\bibinfo {author} {\bibfnamefont {D.~M.}\ \bibnamefont
  {Weinberger}}, \bibinfo {author} {\bibfnamefont {T.~G.}\ \bibnamefont
  {Krause}}, \bibinfo {author} {\bibfnamefont {K.}~\bibnamefont {M{\o}lbak}},
  \bibinfo {author} {\bibfnamefont {A.}~\bibnamefont {Cliff}}, \bibinfo
  {author} {\bibfnamefont {H.}~\bibnamefont {Briem}}, \bibinfo {author}
  {\bibfnamefont {C.}~\bibnamefont {Viboud}}, \ and\ \bibinfo {author}
  {\bibfnamefont {M.}~\bibnamefont {Gottfredsson}},\ }\bibfield  {title}
  {\enquote {\bibinfo {title} {Influenza epidemics in {Iceland} over 9 decades:
  Changes in timing and synchrony with the {United States} and {Europe}},}\
  }\href@noop {} {\bibfield  {journal} {\bibinfo  {journal} {Am. J.
  Epidemiol.}\ }\textbf {\bibinfo {volume} {176}},\ \bibinfo {pages} {649--655}
  (\bibinfo {year} {2012})}\BibitemShut {NoStop}%
\bibitem [{\citenamefont {Housworth}\ and\ \citenamefont
  {Spoon}(1971)}]{Housworth1971}%
  \BibitemOpen
  \bibfield  {author} {\bibinfo {author} {\bibfnamefont {W.~J.}\ \bibnamefont
  {Housworth}}\ and\ \bibinfo {author} {\bibfnamefont {M.~M.}\ \bibnamefont
  {Spoon}},\ }\bibfield  {title} {\enquote {\bibinfo {title} {The age
  distribution of excess mortality during {A2} {Hong-Kong} influenza epidemics
  compared with earlier {A2} outbreaks},}\ }\href@noop {} {\bibfield  {journal}
  {\bibinfo  {journal} {Am. J. Epidemiol.}\ }\textbf {\bibinfo {volume} {94}},\
  \bibinfo {pages} {348--350} (\bibinfo {year} {1971})}\BibitemShut {NoStop}%
\bibitem [{\citenamefont {Bouachache}\ and\ \citenamefont
  {Flandrin}(1985)}]{Flandrin1982}%
  \BibitemOpen
  \bibfield  {author} {\bibinfo {author} {\bibfnamefont {B.}~\bibnamefont
  {Bouachache}}\ and\ \bibinfo {author} {\bibfnamefont {P.}~\bibnamefont
  {Flandrin}},\ }\bibfield  {title} {\enquote {\bibinfo {title} {Wigner-ville
  analysis of time-varying signals},}\ }in\ \href@noop {} {\emph {\bibinfo
  {booktitle} {ICASSP '82. IEEE International Conference on Acoustics, Speech,
  and Signal Processing, Paris}}}\ (\bibinfo {year} {1985})\ pp.\ \bibinfo
  {pages} {1329--1332}\BibitemShut {NoStop}%
\bibitem [{\citenamefont {Flandrin}\ and\ \citenamefont
  {Bernard}(1985)}]{Flandrin1985}%
  \BibitemOpen
  \bibfield  {author} {\bibinfo {author} {\bibfnamefont {P.}~\bibnamefont
  {Flandrin}}\ and\ \bibinfo {author} {\bibfnamefont {E.}~\bibnamefont
  {Bernard}},\ }\bibfield  {title} {\enquote {\bibinfo {title} {Principe et
  mise en oeuvre de l'analyse temps-fr\'equence par transformation de
  wigner-ville},}\ }\href@noop {} {\bibfield  {journal} {\bibinfo  {journal}
  {Traitement du Signal}\ }\textbf {\bibinfo {volume} {2}},\ \bibinfo {pages}
  {143--151} (\bibinfo {year} {1985})}\BibitemShut {NoStop}%
\bibitem [{pyt()}]{pytftb}%
  \BibitemOpen
  \href@noop {} {\enquote {\bibinfo {title} {Python implementation of the
  {TFTB} toolbox developed by {François Auger}, {Olivier Lemoine}, {Paulo
  Gon\c{c}alv{\`e}s} and {Patrick Flandrin}},}\ }\bibinfo {howpublished}
  {https://tftb.readthedocs.io/en/latest/index.html}\BibitemShut {NoStop}%
\bibitem [{\citenamefont {Bestehorn}\ \emph {et~al.}(2022)\citenamefont
  {Bestehorn}, \citenamefont {Michelitsch}, \citenamefont {Collet},
  \citenamefont {Riascos},\ and\ \citenamefont {Nowakowski}}]{Bestehorn2022}%
  \BibitemOpen
  \bibfield  {author} {\bibinfo {author} {\bibfnamefont {M.}~\bibnamefont
  {Bestehorn}}, \bibinfo {author} {\bibfnamefont {T.~M.}\ \bibnamefont
  {Michelitsch}}, \bibinfo {author} {\bibfnamefont {B.~A.}\ \bibnamefont
  {Collet}}, \bibinfo {author} {\bibfnamefont {A.~P.}\ \bibnamefont {Riascos}},
  \ and\ \bibinfo {author} {\bibfnamefont {A.~F.}\ \bibnamefont {Nowakowski}},\
  }\bibfield  {title} {\enquote {\bibinfo {title} {Simple model of epidemic
  dynamics with memory effects},}\ }\href@noop {} {\bibfield  {journal}
  {\bibinfo  {journal} {Physical Review E}\ }\textbf {\bibinfo {volume}
  {105}},\ \bibinfo {pages} {024205--1--10} (\bibinfo {year}
  {2022})}\BibitemShut {NoStop}%
\bibitem [{\citenamefont {Liu}\ and\ \citenamefont
  {Rockl{\"o}v}(2021)}]{Liu2021}%
  \BibitemOpen
  \bibfield  {author} {\bibinfo {author} {\bibfnamefont {Y.}~\bibnamefont
  {Liu}}\ and\ \bibinfo {author} {\bibfnamefont {J.}~\bibnamefont
  {Rockl{\"o}v}},\ }\bibfield  {title} {\enquote {\bibinfo {title} {The
  reproductive number of the delta variant of sars-cov-2 is far higher compared
  to the ancestral sars-cov-2 virus},}\ }\href@noop {} {\bibfield  {journal}
  {\bibinfo  {journal} {Journal of Travel Medicine}\ }\textbf {\bibinfo
  {volume} {28}},\ \bibinfo {pages} {1--3} (\bibinfo {year}
  {2021})}\BibitemShut {NoStop}%
\bibitem [{\citenamefont {Andrews}\ \emph {et~al.}(2022)\citenamefont
  {Andrews}, \citenamefont {Stowe}, \citenamefont {Kirsebom}, \citenamefont
  {Toffa}, \citenamefont {Rickeard}, \citenamefont {Gallagher}, \citenamefont
  {Gower}, \citenamefont {Kall}, \citenamefont {Groves}, \citenamefont
  {O’Connell}, \citenamefont {Simons}, \citenamefont {Blomquist},
  \citenamefont {Zaidi}, \citenamefont {Nash}, \citenamefont {Iwani Binti
  Abdul~Aziz}, \citenamefont {Thelwall}, \citenamefont {Dabrera}, \citenamefont
  {Myers}, \citenamefont {Amirthalingam}, \citenamefont {Gharbia},
  \citenamefont {Barrett}, \citenamefont {Elson}, \citenamefont {Ladhani},
  \citenamefont {Ferguson}, \citenamefont {Zambon}, \citenamefont {Campbell},
  \citenamefont {Brown}, \citenamefont {Hopkins}, \citenamefont {Chand},
  \citenamefont {Ramsay},\ and\ \citenamefont {J.}}]{Andrews2022}%
  \BibitemOpen
  \bibfield  {author} {\bibinfo {author} {\bibfnamefont {N.}~\bibnamefont
  {Andrews}}, \bibinfo {author} {\bibfnamefont {J.}~\bibnamefont {Stowe}},
  \bibinfo {author} {\bibfnamefont {F.}~\bibnamefont {Kirsebom}}, \bibinfo
  {author} {\bibfnamefont {S.}~\bibnamefont {Toffa}}, \bibinfo {author}
  {\bibfnamefont {T.}~\bibnamefont {Rickeard}}, \bibinfo {author}
  {\bibfnamefont {E.}~\bibnamefont {Gallagher}}, \bibinfo {author}
  {\bibfnamefont {C.}~\bibnamefont {Gower}}, \bibinfo {author} {\bibfnamefont
  {M.}~\bibnamefont {Kall}}, \bibinfo {author} {\bibfnamefont {N.}~\bibnamefont
  {Groves}}, \bibinfo {author} {\bibfnamefont {A.-M.}\ \bibnamefont
  {O’Connell}}, \bibinfo {author} {\bibfnamefont {D.}~\bibnamefont {Simons}},
  \bibinfo {author} {\bibfnamefont {P.~B.}\ \bibnamefont {Blomquist}}, \bibinfo
  {author} {\bibfnamefont {A.}~\bibnamefont {Zaidi}}, \bibinfo {author}
  {\bibfnamefont {S.}~\bibnamefont {Nash}}, \bibinfo {author} {\bibfnamefont
  {N.}~\bibnamefont {Iwani Binti Abdul~Aziz}}, \bibinfo {author} {\bibfnamefont
  {S.}~\bibnamefont {Thelwall}}, \bibinfo {author} {\bibfnamefont
  {G.}~\bibnamefont {Dabrera}}, \bibinfo {author} {\bibfnamefont
  {R.}~\bibnamefont {Myers}}, \bibinfo {author} {\bibfnamefont
  {G.}~\bibnamefont {Amirthalingam}}, \bibinfo {author} {\bibfnamefont
  {S.}~\bibnamefont {Gharbia}}, \bibinfo {author} {\bibfnamefont {J.~C.}\
  \bibnamefont {Barrett}}, \bibinfo {author} {\bibfnamefont {R.}~\bibnamefont
  {Elson}}, \bibinfo {author} {\bibfnamefont {S.~N.}\ \bibnamefont {Ladhani}},
  \bibinfo {author} {\bibfnamefont {N.}~\bibnamefont {Ferguson}}, \bibinfo
  {author} {\bibfnamefont {M.}~\bibnamefont {Zambon}}, \bibinfo {author}
  {\bibfnamefont {C.~N.~J.}\ \bibnamefont {Campbell}}, \bibinfo {author}
  {\bibfnamefont {K.}~\bibnamefont {Brown}}, \bibinfo {author} {\bibfnamefont
  {S.}~\bibnamefont {Hopkins}}, \bibinfo {author} {\bibfnamefont
  {M.}~\bibnamefont {Chand}}, \bibinfo {author} {\bibfnamefont
  {M.}~\bibnamefont {Ramsay}}, \ and\ \bibinfo {author} {\bibfnamefont {L.~B.}\
  \bibnamefont {J.}},\ }\bibfield  {title} {\enquote {\bibinfo {title}
  {Covid-19 vaccine effectiveness against the omicron (b.1.1.529) variant},}\
  }\href@noop {} {\bibfield  {journal} {\bibinfo  {journal} {New England
  Journal of Medicine}\ }\textbf {\bibinfo {volume} {386}},\ \bibinfo {pages}
  {1532--46} (\bibinfo {year} {2022})}\BibitemShut {NoStop}%
\bibitem [{\citenamefont {Gostiaux}, \citenamefont {Bos},\ and\ \citenamefont
  {Bertoglio}(2023)}]{Gostiaux2023}%
  \BibitemOpen
  \bibfield  {author} {\bibinfo {author} {\bibfnamefont {L.}~\bibnamefont
  {Gostiaux}}, \bibinfo {author} {\bibfnamefont {W.}~\bibnamefont {Bos}}, \
  and\ \bibinfo {author} {\bibfnamefont {J.}~\bibnamefont {Bertoglio}},\
  }\bibfield  {title} {\enquote {\bibinfo {title} {Periodic epidemic outbursts
  explained by local saturation of clusters},}\ }\href@noop {} {\bibfield
  {journal} {\bibinfo  {journal} {Physical Review E}\ }\textbf {\bibinfo
  {volume} {107}},\ \bibinfo {pages} {L012201--1--7} (\bibinfo {year}
  {2023})}\BibitemShut {NoStop}%
\bibitem [{\citenamefont {Dyson}(2004)}]{Dyson2004}%
  \BibitemOpen
  \bibfield  {author} {\bibinfo {author} {\bibfnamefont {F.}~\bibnamefont
  {Dyson}},\ }\bibfield  {title} {\enquote {\bibinfo {title} {A meeting with
  enrico fermi},}\ }\href@noop {} {\bibfield  {journal} {\bibinfo  {journal}
  {Nature}\ }\textbf {\bibinfo {volume} {427}},\ \bibinfo {pages} {297}
  (\bibinfo {year} {2004})}\BibitemShut {NoStop}%
\end{thebibliography}%

\end{document}